\documentclass[preprint,natbib]{sigplanconf}
\usepackage{stuff,macros-nase}
\conferenceinfo{WXYZ '05}{date, City.} 
\copyrightyear{2008} 
\copyrightdata{[to be supplied]} 

\authorinfo{Andr\'e Hirschowitz}
           {Universit\'e de Nice - Sophia Antipolis}
           {}
\authorinfo{Michel Hirschowitz}
           {CEA - LIST}{}
\authorinfo{Tom Hirschowitz}
           {CNRS, Universit\'e de Savoie}
           {}

 \title{Topological Observations on
   Multiplicative Additive Linear Logic}

\begin{document}
\maketitle

\begin{abstract}
  As an attempt to uncover the topological nature of composition of
  strategies in game semantics, we present a ``topological'' game for
  Multiplicative Additive Linear Logic without propositional
  variables, including cut moves.  We recast the notion of (winning)
  strategy and the question of cut elimination in this context, and
  prove a cut elimination theorem.  Finally, we prove soundness and
  completeness. The topology plays a crucial role, in particular
  through the fact that strategies form a sheaf.
\end{abstract}


\section{Overview}\label{sec:overview}
The notion of a game between two players (P and O) has become
fundamental in proof theory and programming language theory.  A
natural way to think of such a game is as a directed graph, whose
edges represent moves between positions, together with some
information about who plays the moves.

Game semantics~\citep{abramsemint,hyland97} has widened this notion of
game, by providing means to connect two such games together. In game
semantics, each player takes part in two distinct games, and acts as P
in one and as O in the other. Connection, or interaction, then happens
by letting two players respectively play P and O on a common game.

By making several such connections, one obtains a sequence of games,
subject to topological considerations. For example, one may see the
involved games as edges in a graph with the players as vertices, as in
\begin{diagram}[width=1.3cm]
{}  & \rLine^{\textrm{game 0}} & \textrm{player 1} & \rLine^{\textrm{game 1}}
   & \textrm{player 2} & \rLine^{\textrm{game 2}} & \textrm{etc.,}
\end{diagram}
and decree that an open neighborhood of player $i$ is the sequence
\begin{diagram}[width=1.3cm]
 {} & \rLine^{\textrm{game $i - 1$}} & \textrm{player $i$} & \rLine^{\textrm{game
      $i$}} & .
\end{diagram}
The topology here is simplistic, but arguably, this is only due
to the requirement that game semantics be categorical, i.e., each
player sees only two games. This is most striking in the game
semantics of sequent calculi, where sequents $A_1, \ldots, A_n \vdash
B_1, \ldots, B_m$ are interpreted as games $A_1 \wedge \dots \wedge
A_n \to B_1 \vee \dots \vee B_m$.

Let us instead allow each player to see more than two games, i.e., lie
in an open neighborhood like
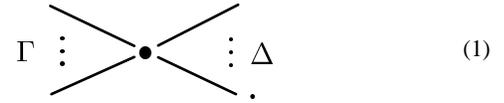
\begin{equation}
\grafflemininline{.5\linewidth}{sq}\label{eq:position}
\end{equation}
We thus consider positions to be spaces obtained by plugging such
atomic neighborhoods together.  A move now leads from a position to
another, where a move -- in the old sense -- has been played on one of
the connections. We investigate this paradigm in the context of
Multiplicative Additive Linear Logic without propositional variables
(henceforth MALL), where logical rules, i.e., moves, are (slightly
enriched) continuous functions between positions. Most emblematic is
perhaps our \emph{cut} move leading from position~(\ref{eq:position})
to
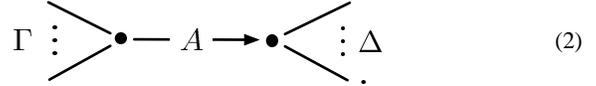
\begin{equation}
  \grafflemininline{.8\linewidth}{pos3}
  \label{eq:position3}
\end{equation}
It is formalised from the obvious continuous function
from~(\ref{eq:position3}) to~(\ref{eq:position}).

We investigate a few topological constructions and properties in this
setting, among which:
\begin{itemize}
\item Strategies, defined in a suitably local way, form a sheaf.
Furthermore, winning strategies are a subsheaf of strategies, i.e.,
the amalgamation of winning strategies is winning again.
\item There is a notion of cut elimination: building upon a
  factorisation system, we define a construction of a cut free
  strategy from a strategy with cuts, again preserving the winning
  character.
\end{itemize}

These observations lead in the case of our semantics for MALL to 
standard logical results like:
\begin{description}
\item[Coherence] There is no winning strategy on the sequent with no
  formula.
\item[Correctness] Any provable MALL sequent admits a winning
  strategy.
\item[Completeness] Any sequent with a winning strategy is provable in
  MALL.
\end{description}

\section{A game for MALL}

\subsection{Hypersequents}\label{subsec:hypersequents}
As explained above, our positions have a particular structure, which
we now define.  First, define MALL formulae by the grammar
$$\begin{array}{rcll}
  \A, \B, \C, \ldots \in \PFormulae & ::= & 
  \zero \alt \un \alt \A \tens \B \alt 
  \A \plus \B \\
  & \alt & \top \alt \bot \alt \A \parr \B \alt \A \with \B,
\end{array}$$
and decree that formulae on the first line are positive, while the
others negative. De Morgan duality is defined as usual (sending a connective
to that vertically opposed to it). Recall in passing the corresponding sequent 
calculus~\citep{ll}.

Say that a \emph{partial} directed graph is a directed graph
\begin{equation}
\begin{diagram}
  E & \pile{\rTo^{s} \\ \rTo_{t}} & V
\end{diagram}\label{eq:hyper}
\end{equation}
with source and target maps $s$ and $t$ \emph{partial}, i.e., edges
may be dangling. We call edges with no source \emph{inputs}, and
dually edges with no target \emph{outputs}.  

\begin{definition}
  A \emph{hypersequent} is a finite, partial directed graph, which is
  furthermore topologically acyclic, i.e., which is acyclic as an
  undirected partial graph.
\end{definition}
Following the intuitions in Section~\ref{sec:overview}, we slightly
abusively identify sequents with connected, one-vertex hypersequents
as in~(\ref{eq:position}).

We then endow hypersequents~(\ref{eq:hyper}) with a topology on the
coproduct $E + V$ by decreeing that a set of points is open when for
each vertex, it contains all the adjacent edges.  Using this topology,
we build a category of hypersquents by defining a morphism $U \to V$
to be given by a continuous function from $U$ to $V$ as
topological spaces, sending vertices to vertices. Such functions
compose in the obvious way.

\begin{remark}[Topology]
  Observe that this entails:
  \begin{itemize}
  \item a set of points is closed iff for each edge it contains all
    the adjacent vertices,
  \item each vertex in $V$ is a closed point,
  \item each edge in $E$ is an open point,
  \item each edge $e \in E$ adjacent to some $v \in V$ has this $v$ in
    its adherence.
  \end{itemize}
\end{remark}

\begin{remark}[Morphisms]
  Morphisms are a bit like morphisms of graphs, in the sense that by
  continuity if an edge $e$ adjacent to some vertex $v$ is sent to an
  edge $e'$, then the image of $v$ is adjacent to $e'$.  However, they
  differ from morphisms of graphs in that:
  \begin{itemize}
  \item they may reverse the direction of edges,
  \item they may sent edges to sequents, as will for example the cut
    move. Such edges are \emph{collapsed} by the morphism, while the other
    are \emph{persistent}.
  \end{itemize}
\end{remark}

To build our category of hypersequents, we define the following
generic way of labeling them. Assume given a category $\cat$ with a
polarity (positive or negative) on morphisms, such that the usual sign
rules are respected by composition, e.g., identities are positive,
composing two negative morphisms yields a positive one, etc.  Define
the category $\hype{\cat}$ of $\cat$-hypersequents to have
\begin{itemize}\item objects:
  hypersequents with edges labeled in $\Ob{\cat}$, i.e., equipped with
  a function $\ell: E \to \Ob{\cat}$;
\item morphisms $U \to V$: pairs $(g, o)$ of a morphism $g: U \to V$
  of unlabeled hypersequents, and for each persistent edge $e$, a
  morphism $o_e: \ell_U (e) \to \ell_V (g (e))$ in $\cat$, such that
  if $o_e$ is positive then the direction of $e$ is preserved by $g$,
  and otherwise it is reversed\footnote{Here by direction we mean the
    pair $(se, te)$, seeing $s$ and $t$ as functions $E \to (V +
    1)$. An edge without source or target may thus have its direction
    both preserved and reversed.}.
\end{itemize}
Morphisms compose, and the condition on the direction of edges is
preserved thanks to the sign rules.

We apply this construction to the category $\Coor$ with objects the
positive formulae and morphisms $\A \to \B$ the \emph{occurrences},
i.e., paths from the root in $\B$ reaching a subformula equal to $\A$
up to de Morgan duality. The sign of a morphism is that of the
subformula reached by the path.  This gives us the category $\GH =
\hype{\Coor}$.

Before going on to define the moves of our game, we show a few example
morphisms. From the obvious continuous function from (the underlying
space of)
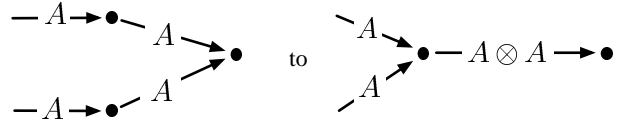
\begin{center}
  \grafflemininline{.4\linewidth}{extens1} \hfil to \hfil
  \grafflemininline{.45\linewidth}{extens2}
\end{center}
we may define four different morphisms, according to the occurrences
we assign to the two premises of the tensor. For example, we may send both
edges to the first premise by assigning them both the occurrence $0$.
We also may assign the upper edge the occurrence $1$ and to the 
lower edge the occurrence $0$. There are two symmetric morphisms.

To illustrate the conventions on signs of formulae, consider the
morphism from 
\begin{center}
  \begin{tabular}{c}
    \graffleinline{orient1}\ \ \ \ {} \\ to \\
    \graffleinline{orient2}
  \end{tabular}
\end{center}
It assigns occurrence $10$ to the unique edge of the domain. But since
the corresponding subformula of $\A \tens ((\B \parr \C) \parr D)$ is
negative, the edge's source and target are swapped, and the formula is
dualised.  Of course, we immediately introduce the notation consisting
of labeling edges with negative formulae to denote the reversed edge
with the dual formula. In this way, the domain of the above morphism
becomes \grafflecenter{orient3} (We could also have used an equivalent
category where labels may directly be negative.)

\subsection{Moves}\label{subsec:moves}
In the category $\GH$ of hypersequents, we now single out a class of
morphisms as our \emph{\propermoves,} thus forming a subgraph $\MA$ of
$\GH$.  We will first define a set of \emph{basic moves} corresponding
to the rules of MALL, and then extend them by embedding.
\begin{figure}[t]
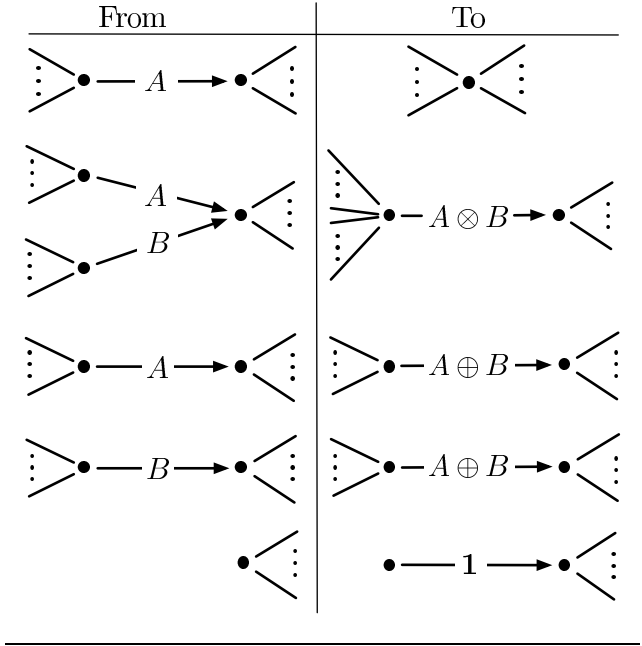
 
  \grafflecenter{basicmoves}{}
  \caption{Basic moves}
  \label{fig:basic:moves}
\end{figure}

Our basic moves are defined in Figure~\ref{fig:basic:moves}. Each line
defines a move, the first being the already mentioned \emph{cut} move.
In each case, the move is the obvious morphism from left to right, the
dots meaning that the move is a morphism on a larger hypersequent,
which is an isomorphism outside the shown part.

Since we want to get topological, it seems natural to consider
restrictions of basic moves.  For example, the restriction of the
tensor move to the left-hand sequent would send
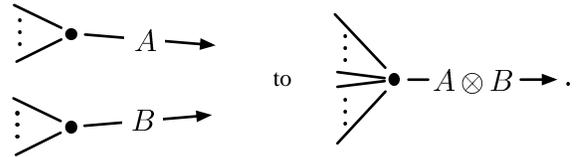
\begin{center}
  \grafflemininline{.35\linewidth}{tens6} \hfil to \hfil
  \grafflemininline{.4\linewidth}{tens5}
\end{center}
To formalise this idea, we consider the identity-on-objects
subcategory $\HH \rInto \GH$ with the same objects, and morphisms the
pairs $(g, o)$ with $g$ an open embedding and $o$ the function
assigning to each edge labeled $\A$ the identity occurrence
$\id_\A$. In the following, we call these morphisms simply
\emph{embeddings}. Observe that $\GH$ has pullbacks along embeddings,
that pullbacks of embeddings are embeddings again.

We can now extend our basic moves under the following rule: if a
morphism $m$ as above is the restriction of a basic move $m'$ along an
embedding $j$, in a pullback square
\begin{equation}
  \begin{diagram}
    U \SEpbk & \rInto^i & U' \\
    \dTo<{m} & &  \dTo>{m'} \\
    V & \rInto^j & V',
  \end{diagram}\label{eq:restmove}
\end{equation}
and if further $m$ is not an isomorphism, then $m$ is a \emph{\propermove.}

Finally, a vertex $v$, is \emph{active} in a \propermove $m$ when
either
\begin{itemize}
\item $m$ is a cut and $v$ is the cut sequent, or
\item $m$ is not a cut and $v$ is the source of the broken edge.
\end{itemize}
There is at most one active vertex in a \propermove, and we call
sequents and \propermoves active when they contain an active vertex,
and \emph{passive} otherwise.

\subsection{Plays and strategies}\label{subsec:axioms}
To sum up, we have a site $\GA$ of hypersequents, with
\begin{itemize}
\item an identity-on-objects subcategory $\HA \rInto \GA$ of
  embeddings,
\item an identity-on-objects subgraph $\MA \rInto \GA$ of \propermoves,
stable under composition with isomorphisms,
\end{itemize}
such that
\begin{itemize}
\item embeddings have pullbacks in $\GA$, and these pullbacks are
  embeddings again,
\item the pullbacks of \propermoves along embeddings thus exist, and
  are either \propermoves again, or isomorphisms.
\end{itemize}
We also have a polarity on \propermoves, i.e., a partition of \propermoves into
passive and active ones.

Let us now define plays in this setting.  Traditionally, plays are
defined as sequences of moves. Here, because of the topological nature
of positions, we find it useful to generalise this as follows.
Consider the graph $\MGA$ of \emph{\generalizedmoves }defined by the
following pushout of graphs
\begin{diagram}
  \Ob{\GH} & \rInto & \HH \\
  \dInto &  & \dInto \\
  \MA & \rInto & \NWpbk \MGA.
\end{diagram}
It has
\begin{itemize}
\item vertices the objects of $\GA$, and
\item edges the coproduct of \propermoves and embeddings.
\end{itemize}
A \emph{ play} on some object $U$ is a path to $U$ in $\MGA$; it is
\emph{proper} when it has no embeddings. Let $\PA$ be the free
category generated by $\MGA$. Composition defines a functor $\geom:
\PA \to \GA$, which leave implicit except where necessary.

Let us now turn to strategies. Traditionally, strategies are
non-empty, prefix-closed sets of (proper) plays. Here, we are in a
topological setting, so instead of defining strategies as sets of
plays, we want to include in them as local an information as
possible. What strategies have to contain is, at each stage in the
course of the play, for each involved edge or sequent, the moves it
accepts.  We formally define them to contain this information and not
more. Still, (winning) strategies generate meaningful sets of plays,
as we explain in a bit more detail in Section~\ref{subsec:winning}.

Call a hypersequent \emph{atomic} when it is either empty, or an edge,
or a sequent. A \emph{thread} on a hypersequent $U$ is a play $\play$
such that:
\begin{quote}
  (T) For all \propermoves $m: W \to V$ appearing in $\play$, $V$ is atomic.
\end{quote}
Now, call a \generalizedmove $W \rTo^{f}
V$ \emph{mandatory} when either
\begin{itemize}
\item $f$ is an embedding, or
\item $V$ is atomic and $f$ is a passive \propermove.
\end{itemize} A \emph{strategy} on $U$ is then a set of threads $\proc$
which is:
\begin{description}
\item[S1] prefix-closed, i.e., if $t t' \in \proc$, then also $t \in
  \proc$,
\item[S2] stable under extension by mandatory \generalizedmoves, i.e., if
  $W \rTo^{f} V$ is mandatory and $V \rTo^{t} U$ is in $\proc$, then
  also $tf$ is in $\proc$,
\item[S3] stable under isomorphism, i.e., if for any threads
  \begin{center}
    \hfil $t': U \to X$ \hfil and \hfil $t: Y \to V$, \hfil
  \end{center}
  and commuting square
  \begin{diagram}[height=.5cm,width=.7cm]
    X & \rTo^{j} & X' \\
    \dTo<{m} & & \dTo>{m'} \\
    Y & \lTo^{i} & Y'
  \end{diagram} with $m$ and $m'$ \generalizedmoves and $i$ and $j$
  isomorphisms, $t m t' \in \proc$ iff $t i m' j t' \in \proc$;
\item[S4] and stable under composition and decomposition of
  embeddings, i.e., for any $t, t'$ as above and any embeddings
  \begin{diagram}
    X & \rInto^{h'} & Z & \rInto^{h} & Y,
  \end{diagram}
  $t \rond h \rond h' \rond t' \in \proc$ iff $t \rond \geom(hh')
  \rond t' \in \proc$
\end{description}
Observe that these axioms entail the ``one-sided'' versions of
\textbf{S3}: if, e.g., $i$ is the identity, then $t m t' \in \proc$
iff $t m' j t' \in \proc$. Indeed, we apply \textbf{S3} twice
with the squares
  \begin{diagram}[height=.5cm,width=.7cm]
    X & \rTo^{j} & X'  & \rEqto & X'\\
    \dTo<{m} & & \dTo>{m'} & & \dTo>{m'} \\
    Y & \lEqto & Y & \rEqto & Y
  \end{diagram}
  to deduce that $t \rond \id \rond \id \rond m' \rond \id \rond j
  \rond t'$ is in $\proc$, then apply \textbf{S4} with $\id \rond
  \id$, and apply \textbf{S3} again with the right-hand square above,
  to obtain that $t m' j t'$ is in $\proc$. The converse implication
  is similar.

The \emph{restriction} $t^* (\proc)$ of a set $\proc$ of threads on
$U$ along some thread $t: V \to U$ is the set of threads $t'$ on $V$
such that $t t' \in \proc$.  The restriction $t^* (\proc)$ of a
strategy along any $t$ is obviously a strategy again, although
possibly the empty one. (Observe in passing that a strategy may be
empty.)  We furthermore have, thanks to \textbf{S4}, for the obvious
Grothendieck topology on $\HH$,
\begin{theorem}
  Strategies form a sheaf $\stratsH: \op{\HH} \to \Set$.
\end{theorem}
\begin{proof}
  A strategy $\proc$ on $U$ is determined by its set of restrictions
  $h^* (\proc)$ for $h: V \rInto U$ and $V$ atomic. But any covering
  sieve on $U$ includes those $h$'s and thus entirely determines
  $\proc$.  So the presheaf $\stratsH$ is separated (amalgamations,
  when they exist, are unique).  Now given a sieve $\sieve$ on $U$
  with compatible strategies $\proc_i$ on the embeddings $h_i: U_i
  \rInto U$ of $\sieve$, define the amalgamation $P$ as follows.
  First any sequence of embeddings to $U$ is in $P$. Furthermore, for
  any thread $p$ on $U$ decomposing as
  \begin{diagram}
    W & \rTo^{q} & V & \rInto^{r} & U
  \end{diagram}
  with $r$ a sequence of embeddings and $V$ atomic, then let $U_i$ be
  one of the members of $\sieve$ isomorphic to $V$, and $q': W \to
  U_i$ be the play corresponding to $q$ there (recall that \propermoves are
  stable under isomorphism).  Then decree that $p \in P$ iff $q' \in
  \proc_i$.
\end{proof}

\section{Cut elimination}
In this section, we define our cut elimination (= descent) procedure
for strategies.  We start by specifying cut elimination as a function
from strategies to cut free strategies: Consider the subsheaf $\cfH
\rInto \stratsH$ of strategies consisting of cut free strategies,
i.e., those whose plays are all in the free category $\PNCH$ generated
by non cut \generalizedmoves.  Cut elimination should provide a
morphism of sheaves $\cutelimfun: \stratsH \to \cfH$, preserving the
winning character of strategies. In this section we stick to defining
our morphism of sheaves, and defer to Section~\ref{sec:logic} the
study of winning strategies.

\subsection{Overview}\label{subsec:celim:overview}
\paragraph{Remote view: an easy task}
We will construct our morphism of sheaves using a more general family
of functions $\descendcfun: \stratsH (U) \to \cfH (V)$ indexed by a
particular class of morphisms $U \rTo^{c} V$ in $\GH$. The
subfamily of functions $\stratsH (U) \to \cfH (U)$ obtained by taking
$c = \id_U$ will lead to the desired morphism of sheaves.  

The involved class of morphisms $c$ is that of \emph{cut only
  topological plays}, i.e., morphisms $c$ as above admitting a
decomposition into cut moves. For each such morphism, we will define
functions $\descendcfun: \stratsH (U) \to \cfH (V)$ sending strategies
on $U$ to cut free strategies.

The rough idea for defining these functions is natural: compute cut
elimination for \generalizedmoves and extend it to plays by
induction. Cut elimination for moves arises from a factorisation
system: a morphism in $\GH$ may always be decomposed into a
``cut-like'' morphism, followed by a ``non cut-like'' morphism, which
yields a factorisation system $(\LLH, \RRH)$. In particular, cut only
topological plays are in $\LLH$, but $\LLH$ contains other morphisms,
as we shall shortly see.

Given a \generalizedmove $W \rTo^{m} U$, factorisation yields
the dashed arrows in
\begin{equation}
\begin{diagram}
  W  & \rDashto^{c'} & X\\
\dTo<{m} & & \dDashto>{q} \\
U & \rTo^{c} & V,
\end{diagram}\label{eq:descent:move}
\end{equation}
with $c' \in \LLH$ cumulating the cuts in $c$ and $m$, and $q \in
\RRH$.  We thus take $\descendc{m} = q$, and say that $m$
\emph{descends} along $c$ as $q$.

It turns out that there are (roughly) two relevant configurations
here:
\begin{itemize}
\item $q$ is a \generalizedmove, or
\item $q$ is an identity.
\end{itemize}
We interpret the second case by saying that $\descendcfun$ should really
send \generalizedmoves to \emph{plays}. If $q$ is a \generalizedmove,
then $\descendc{m}$ is the one-move play $q$. Otherwise, $\descendc{m}$ is
the empty play, and we replace the above square by a triangle
\begin{diagram}
  W  \\
  \dTo<{m} & \rdDashto>{c'} \\
  U & \rTo^{c} & V.
\end{diagram}

Cut elimination (= descent) for plays is then obtained by piling such
squares and triangles: given a play $W \rTo^{p} U$, this yields the
dashed arrows in
\begin{equation}
\begin{diagram}
  W  & \rDashto^{c'} & X\\
\dTo<{p} & & \dDashto>{q'} \\
U & \rTo^{c} & V,
\end{diagram}\label{eq:play:above:c}
\end{equation}
where $q'$ is the concatenation of the plays obtained as above, for
each move of $p$.

\paragraph{Cut elimination is partial}
This should define descent for plays, but things turn out to be a
little more complicated, because the function $\descendcfun$ is actually
\emph{partial}. Indeed, some embeddings cause trouble, as shown by the
following example. The following is a factorisation square: 
\begin{diagram}[tight,width=.3\linewidth,height=1.3cm,shortfall=.3cm]
  {\grafflemininlinecenter{.4\linewidth}{fail3}} & \rTo^{c'} &
  {\grafflemininlinecenter{.2\linewidth}{fail4}} \\
  \dInto<{h} & & \dTo>{h'} \\
  {\grafflemininlinecenter{.4\linewidth}{fail1}} & \rTo^{c} &
  {\grafflemininlinecenter{.2\linewidth}{fail2}}
\end{diagram}
Indeed, the lower-left composite may be decomposed as 
\begin{itemize}
\item a collapse of the $\A$-labeled edge,
\item an injection of the resulting vertex into the codomain.
\end{itemize}
We cannot consider this a successful descent, for two reasons:
\begin{itemize}
\item $c'$, although in $\LLH$, is no cut only topological play --
  cuts only collapse two-ended edges, and
\item $h'$, although in $\RRH$, is not open -- since its image is not.
\end{itemize}
So our function is partial. Worse, it is much likely to be undefined
on threads, which very often behave as $h$ above, i.e., restrict to
one end of an edge collapsed by $c$. Even worse, threads may restrict
to one end of an edge created by some cut move earlier in the thread.

We thus cannot reasonably define descent for strategies as a
direct extension of descent for plays, i.e., by taking
$\descendc{\proc}$ to be the image of $\proc$ under $\celimfun$ (for
some strategy $\proc$).

But, we may delineate the problem better: in a play $p$ as
in~(\ref{eq:play:above:c}), call an edge \emph{doomed} when it is
collapsed by the composite $cp$ (we indeed want doomed edges to
disappear through cut elimination). At such a stage $p$, observe that
partiality is only caused by embeddings cutting off doomed edges. 
We thus adapt the notion of thread to a context where doomed edges are
considered unbreakable. This leads to the following notion of
$c$-\emph{cable}.

\paragraph{Characterising the plays descending to threads}
Before playing, threads restrict to atomic hypersequents, which we now
view as connected subspaces with no two-ended edge. If we now
consider, at each stage, doomed edges as unbreakable, the atomic
hypersequents should now be the connected hypersequents
where \begin{center} doomed $\Leftrightarrow$ two-ended,
\end{center}
i.e., a doomed arrow has two ends, and a non-doomed arrow does not.
Our $c$-cables are thus the plays which, before playing a \propermove,
restrict to such subspaces. In short: before playing, cables must cut
off all the edges that may be cut off.  Observe that if there are no
doomed edges, one exactly recovers threads.  

We then may define the descent $\celimc{\proc}$ of a strategy $\proc$
to be $\descendc{\pipesofc{\proc}}$, i.e., the image by $\descendcfun$ of
its $c$-cables.

\paragraph{Finalisation}
Then we are almost done. Beyond being partial, our function
$\descendcfun$ was actually only defined up to isomorphism, as is
factorisation.  We thus define it as a relation, but the construction
remains essentially the same.  Finally, the cut elimination of $\proc$
is a set of threads, but need not be a strategy, and we need to close
off by Axiom \textbf{S4} to obtain one. To explain why this is so,
recall that Axiom \textbf{S4} requires strategies to contain plays
regardless of composition of embeddings, i.e., $\geom(h \rond h')$ is
not distinguished from the sequence $h \rond h'$.  Now, consider a
descent like
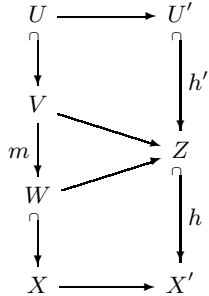
\begin{diagram}[height=.3cm]
  U & \rTo & U' \\
    & &    \\
 \dInto   & &     \\
    & &   \dInto>{h'} \\
  V & &    \\
    & \rdTo &    \\
  \dTo<{m}  & & Z  \\
    & \ruTo &    \\
  W & &    \\
    & &   \dInto>{h} \\
 \dInto  & &    \\
    & &    \\
X   & \rTo & X'
\end{diagram}
as above, where some \propermove $m$ descends to the identity.  The
composite $\geom(h \rond h')$ need not be the image of any play on
$X$. The other direction of \textbf{S4} is satisfied though, so we
need only close under composition of embeddings, defining the descent
of a strategy $\proc$ to be the corresponding closure
$\cutelimc{\proc} = \closure{\celimc{\proc}}$.

\subsection{Factorisation}
\begin{figure}[t]
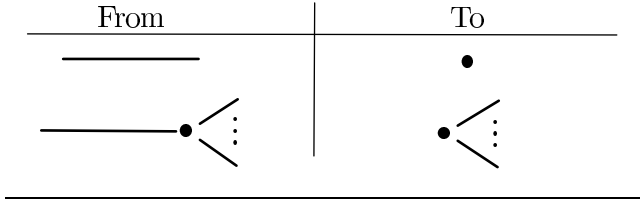

  \grafflecenter{newmoves}{}
  \caption{New steps}
  \label{fig:new:moves}
\end{figure}

Let us start with the announced factorisation system.  Given a
morphism $U \rTo^{f} V$, we may decompose it as
\begin{diagram}
  U & \rTo^{g} & W & \rTo^{h} & V,
\end{diagram}
where all $g$ does is collapse edges to vertices. Formally, $g$
belongs to the class $\LLH$ of morphisms which may be decomposed into
a sequence using only cut moves and the morphisms shown in
Figure~\ref{fig:new:moves}, plus isomorphisms.

Now, what will $h$ look like? Obviously, $h$ will send edges to edges.
And this turns out to be enough: calling $\RRH$ the class of morphisms
sending edges to edges, we have

\begin{lemma}\label{lemma:factoVS}
  The classes $\LLH$ and $\RRH$ form a factorisation system for
  $\GH$.
\end{lemma}

To prove this, we first observe that morphisms $U \rTo^{c} V$ in
$\LLH$ are epi.  Indeed, their underlying functions are surjective,
and moreover, for edges of $V$, precomposition by $c$ does not change
the occurrences. This leads to
\begin{proof}[Proof of Lemma~\ref{lemma:factoVS}]
  Existence of a factorisation is obvious. Now,
  consider a commuting square
  \begin{equation}
  \begin{diagram}
    X & \rTo^{f} & Y \\
    \dTo<{c} & & \dTo>{r} \\
    U & \rTo^{g} & V
  \end{diagram}\label{eq:square:facto}
\end{equation}
  with $c \in \LLH$ and $r \in \RRH$.
  Then choose a factorisation $(c', r')$ for $f$, as in
  \begin{diagram}[height=.3cm]
    & & W & & \\
    & \ruTo<{c'} & & \rdTo>{r'} & \\
    X && \rTo_{f} && Y.
  \end{diagram}

  Now, for any morphism $X \rTo^{f} Y$, let $\collapsed{f}$ be the set
  of edges in $X$ collapsed by $f$, i.e., sent to vertices.

  We have $$\collapsed{c} \subseteq \collapsed{f} = \collapsed{c'}.$$
  By collapsing exactly the edges in $c (\collapsed{c'})$, we define
  a morphism $c''$ such that 
  \begin{diagram}[height=.3cm]
    & & U & & \\
    & \ruTo<{c} & & \rdTo>{c''} & \\
    X && \rTo_{c'} && W.
  \end{diagram}
  All in all, we obtain a diagram
  \begin{diagram}
    X & & \rTo^{f} & & Y \\
      & \rdTo>{c'} & & \ruTo<{r'} \\
      \dTo<{c} && W && \dTo>{r} \\
      & \ruTo>{c''} & & \mbox{?} \\
      U && \rTo^{g} && V,
  \end{diagram}
  where the upper triangles and the perimeter are known to commute.
  But a simple diagram chase shows that $c$ equalises the lower
  triangle, i.e., $gc = r r' c'' c$. But $c$ is epi, so the lower
  triangle commutes.

  This yields a diagonal for the original
  square~(\ref{eq:square:facto}), making both triangles commute.  Its
  uniqueness is a direct consequence of $c$ being epi.
\end{proof}

\subsection{The partial ``function''}
We then define our relation on plays (which is more like a partial
function up to isomorphism), defined as a bipartite graph $\descend$:
when is a cut free play the cut elimination of a given play?  In order
to define this, we start with the corresponding relation on
\generalizedmoves, not trying for the moment to understand which moves
have an image. Consider the graph $\MCA$ with vertices the cut only
topological plays $c: U \to V$, and two kinds of edges $c \to c'$,
based on the squares
\begin{center}
  \begin{tabular}{*{3}{p{.27\linewidth}}}
    \begin{minipage}[c]{\linewidth}
      \begin{diagram}
        U'  & \rTo^{c'} & V' \\
        \dTo>{m} & & \dTo<{m'} \\
        U & \rTo^{c} & V
      \end{diagram}
    \end{minipage}
    &
    \begin{minipage}[c]{\linewidth}
      \begin{diagram}
        U'  & \rTo^{c'} & V' \\
        \dInto>{h} & & \dInto<{h'} \\
        U & \rTo^{c} & V
      \end{diagram}
    \end{minipage}
    &
    \begin{minipage}[c]{\linewidth}
      \begin{diagram}
        U'  & \rTo^{c'} & V' \\
        \dTo>{m} & & \dEqto \\
        U & \rTo^{c} & V,
      \end{diagram}
    \end{minipage}
  \end{tabular}
\end{center}
where
\begin{itemize}\item 
  the right-up sequence is a $(\LLH, \RRH)$-factorisation of the
  left-low composite,
\item $m$ and $m'$ are \propermoves, with $m'$ non cut,
\item $h$ and $h'$ are embeddings,
\item $c$ and $c'$ are cut only topological plays.
\end{itemize}
We define our graph $\MCA$ to have as edges $c \to c'$ the squares as
the first two above, and the triangle
\begin{diagram}
      U' &  & \\
      \dTo<{m} & \rdTo^{c'} & \\
      U & \rTo^{c} & V  
\end{diagram}
for each square as the third one above.  This graph freely generates a
category $\PCH$ whose morphisms $c \to c'$ are piles of such squares
and triangles.  Taking the left- and right-hand sides of such piles
yields source and target functors to the category $\PH$ of
plays. However, we make the distinction with the category $\PNCH$ of
cut free plays, and denote by
\begin{diagram}
  \PH & \lTo^{s} & \PCH & \rTo^{t} & \PNCH
\end{diagram}
the corresponding source and target functors. This defines a bipartite
graph $\descend$ between plays and cut free plays, and we say that $p$
\emph{descends} along $c$ as $q$ when there is an edge $p \to q$ in
$\descend$ with lower border $c$.  

This notion of descent extends by union to a function on sets of plays
$S$: $\descend_c{S}$ is the set of plays descending from plays in $S$
along $c$.

However, this does not meaningfully send strategies to strategies,
because threads do not in general descend along such $c$'s. Nor can we
prove for free that it preserves the winning character: the image of a
given strategy could \emph{a priori} be empty.

So, in the next section, we start investigating conditions for
\generalizedmoves and plays to descend along a given $c$ in
Section~\ref{subsec:descending:moves}. This leads to the notion of
\compatibility{$c$}.  Using this, we define our $c$-cables in
Section~\ref{subsec:cables}, which all descend to threads along $c$.
We then turn back to strategies, and after defining
(Section~\ref{subsec:threads:of:play}) the plays $\playsof{\proc}$
generated by a strategy $\proc$ over $V$, we define the descent of
$\proc$ along $c$ to be the set of threads over $V$ descending from a
play in $\playsof{\proc}$. However, the result of descent need not be
a strategy, and we still must close under (one direction of) axiom
\textbf{S4}. We show in Section~\ref{sec:logic} that this notion of
descent preserves the winning character (as defined there) of
strategies. Cut elimination is recovered as the special case where $c$
is the identity.

\subsection{When plays descend: compatibility}\label{subsec:descending:moves}
We now turn to characterising moves that descend along a given $c$.
We further give a sufficient condition for descending plays:
\compatibility{$c$}.

\paragraph{Characterising cut only topological plays}
Recall that a \emph{topological play} is any map in $\GH$ which may be
decomposed into \generalizedmoves. Further call \emph{cut only} any
morphism in $\LLH$. A cut only morphism does not have to be a
topological play, as shown for example by the morphisms in
Figure~\ref{fig:new:moves}.  Indeed, cut moves only collapse
two-ended edges.  However, given any morphism $c: U \to V$, the
following are easily shown equivalent:

\begin{description}
\item[\textit{(i)} ] $c$ is a topological play and a cut only
  morphism,
\item[\textit{(ii)}] $c$ is cut only, and it only collapses two-ended edges,
\item[\textit{(iii)}] $c$ admits a decomposition into cut moves.
\end{description}
Hence, it is consistent to take as we did \emph{cut only topological
  play} to mean topological play admitting a decomposition into cut
moves.

\paragraph{\Propermoves}
We start by proving that everything goes smoothly for \propermoves.
\begin{lemma}\label{lemma:non:dangling}
  If $U \rTo^{m} V$ is a \propermove and $e$ is a two-ended edge in $V$,
  then the edges in $\inv{m} (e)$ are also two-ended in $U$.
\end{lemma}
\begin{proof}
  By case inspection this holds for basic moves, and it remains true
  after any restriction.
\end{proof}

Recall that $\collapsed{p}$ denotes the set of edges in $U$ collapsed
by $p$, i.e., sent to vertices, for any $U \rTo^{p} V$.

\begin{lemma}\label{lemma:descent}
  Any sequence
  \begin{diagram}[inline,width=.7cm]
    W & \rTo^{m} & V & \rTo^{c} & U
  \end{diagram}
  with $m$ any \propermove and $c$ a cut only topological play may be
  completed as a commuting square
  \begin{equation}
  \begin{diagram}[height=.6cm]
    W & \rTo^{c'} & W' \\
    \dTo<{m} & & \dTo>{f} \\
    V & \rTo^{c} &  U
  \end{diagram}\label{eq:square:descent}
\end{equation}
in $\GH$, with $f$ an isomorphism or a cut free \propermove and $c'$ a
cut only topological play. Furthermore, $\collapsed{cm} =
\collapsed{c'}$.
\end{lemma}

\begin{proof}
  Let $E^c_V$ be the set of edges collapsed by $c$, and $E^c_W$ their
  antecedents by $m$.  By Lemma~\ref{lemma:non:dangling}, the edges in
  $E^c_W$ are two-ended.

  If $m$ is a cut move, then let $E^m_W$ be the set of edges collapsed
  by $m$. We have by construction $E^c_W$ and $E^m_W$ disjoint. Let
  $c': W \to W'$ be the cut only map obtained by collapsing exactly
  these edges $E_W = (E^c_W \uplus E^m_W)$ in $W$. Since the edges in
  $E_W$ have two ends, $c'$ is a topological play, and we have an
  isomorphism $f: W' \iso U$.

  If $m$ is not a cut move, let $c': W \to W'$ be the cut only
  topological play collapsing exactly $E^c_W$. It remains to find $f$
  as in~(\ref{eq:square:descent}). For this, let $E^m_W$ be the set of
  edges in $W$ which are not assigned the empty occurrence by $m$,
  i.e., which are acted upon by $m$. All such edges are sent to a
  unique edge $e_0$ in $V$, and being in $E^c_W$ for $e \in E^m_W$ is
  the same as being in $E^c_V$ for $e_0$. Thus, we have either $E^m_W
  \subseteq E^c_W$, or $E^m_W$ disjoint from $E^c_W$.

  Now, if $E^m_W \subseteq E^c_W$, we again have an isomorphism $f: W'
  \iso U$, and we are done. Otherwise, $E^m_W$ is left untouched by
  $c'$, and we may mimic the action of $m$ on the image of $E^m_W$ by
  $c'$, and land in $U$, making the square~(\ref{eq:square:descent})
  commute in $\GH$.

  In all cases, clearly, $\collapsed{cm} = \collapsed{c'}$.
\end{proof}

\paragraph{Embeddings}
We have seen in Section~\ref{subsec:celim:overview} that this does not
work for embeddings in general.  However, the process works smoothly
when such an $h$ does not cut off any edge collapsed by $c$. Formally:
\begin{lemma}\label{lemma:descent:embedding}
For any square
\begin{equation}
    \begin{diagram}
      U'  & \rTo^{c'} & V' \\
      \dInto<{h} & & \dTo>{h'} \\
      U & \rTo^{c} & V
    \end{diagram}\label{eq:descent:embedding}
  \end{equation}
    with $h$ an embedding, $c$ a cut only topological play,
    and $(c', h')$ an $(\LLH, \RRH)$-factorisation of $ch$,
    the following are equivalent
    \begin{description}
    \item[\textit{(i)}] $h'$ is an embedding and $c'$ is a cut only
      topological play, 
    \item[\textit{(ii)}] any edge $e \in \collapsed{ch}$ has two ends
      in $U'$,
    \item[\textit{(iii)}] any two-ended edge $e \in \collapsed{c}$ in
      the image of $h$ has a two-ended antecedent in $U'$.
    \end{description}
    Furthermore, in this case $\collapsed{ch} = \collapsed{c'}$ and
    the square is a pullback.
\end{lemma}
We first prove two easy lemmas:
\begin{lemma}\label{lemma:one:all}
  Consider a diagram
  \begin{diagram}[inline,width=.7cm]
    U' & \rInto^{h} & U & \rTo^{c} & V
  \end{diagram}
  with $h$ an embedding, $c$ a cut only topological play, and such
  that any edge $e \in \collapsed{ch}$ has two ends in $U'$.  For any
  vertex $v_U \in U$ in the image of $h$, all of $\inv{c} (c (v_U))$
  is in the image of $h$ too.
\end{lemma}
\begin{proof}
  Let $v_V = c (v_U)$. Since $c$ is a play, $\inv{c} (v_V)$ is
  connected and has only two-ended edges, hence is a tree in the
  graph-theoretical sense. But $h$ is open, so any edge $e$ adjacent
  to $v_U$ has an antecedent by $h$.  But since any edge $e \in
  \collapsed{ch}$ has two ends in $U'$, the other end $v'$ of $e$ also
  has an antecedent. But similarly any edge incident to $v'$ has an
  antecedent. By induction on the length of the path from $v_U$, all
  of $\inv{c} (v_V)$ is in the image of $h$.
\end{proof}

Here is the second lemma:
\begin{lemma}\label{lemma:decomp:embed}
  If in a triangle
  
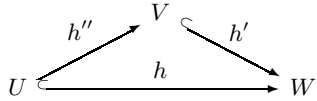
\begin{diagram}[height=.5cm]
     & & V & & \\
     & \ruTo<{h''} & & \rdInto>{h'} \\
   U & & \rInto^{h} & & W
 \end{diagram}
 $h$ and $h'$ are embeddings, then so is $h''$.
\end{lemma}
\begin{proof}
  Obviously, $h''$ is injective since $h$ is. Moreover, since $h$ and
  $h'$ have empty occurrences, $h''$ has empty occurrences. Finally,
  $h''$ is open: for any open $X \subseteq U$, $h'' (X)$ is equal to
  $\inv{h'} (h (X))$, which is open since $h'$ is continuous and $h$
  is open.
\end{proof}

We turn back to the proof of Lemma~\ref{lemma:descent:embedding}.
\begin{proof}[of Lemma~\ref{lemma:descent:embedding}.]
  First of all, $\collapsed{ch} = \collapsed{h'c'} = \collapsed{c'}$.

  Then, \textit{(ii)} implies \textit{(iii)}, because any $e \in
  \collapsed{c}$ in the image of $h$ is in $\collapsed{ch}$.
  
  Conversely, \textit{(iii)} implies \textit{(ii)}, because given $e
  \in \collapsed{ch}$, $h (e)$ is in $\collapsed{c}$. But $c$ is a
  play, so $h (e)$ has two ends, and so by \textit{(iii)}, $e$ too has
  two ends.

  Furthermore, \textit{(i)} implies \textit{(ii)}, since $c'$ is a
  play.

  Finally, if any edge $e \in \collapsed{ch}$ has two ends in $U'$,
  since $c'$ collapses exactly the doomed edges in $U'$ and these are
  all two-ended, $c'$ is a cut only topological play.

  To show that $h'$ is open, first consider any vertex $v_V$ in $V$,
  and any edge $e_V$ incident to $v_V$. There is a unique pair $(v_U,
  e_U)$ with $e_U$ incident to $v_U$ in $U$, sent to $(v_V, e_V)$ by
  $c$.  (Indeed, $c$ leaves persistent edges untouched and does not
  augment their adherence.)

  Now, if $v_V$ is in the image of $h'$, then it is in the image of
  $c'h'$, because $c'$ is surjective. Moreover, since $c'$ is a play,
  it has some antecedent vertex $v_{U'}$ in $U'$. Now, let $v'_{U} = h
  (v_{U'})$.  It is sent to $v_V$ by $c$, so by
  Lemma~\ref{lemma:one:all}, all of $\inv{c} (v_V)$ is in the image of
  $h$. Hence $v_U$ has an antecedent $v'_{U'}$ in $U'$. But since $h$ is
  open, $e_U$ also has an antecedent, left untouched by $c'$, and hence
  $e$ has an antecedent by $h'$.  So, $h'$ is open and \textit{(ii)}
  implies \textit{(i)}.

  Finally, consider the morphism $f$ induced by universal property
  of pullback in
  
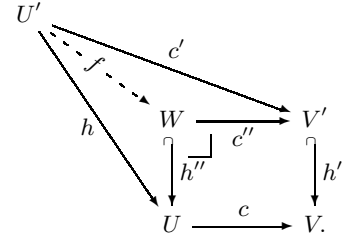
\begin{diagram}
    U'  \\
      & \rdTo (4,2)>{c'} \rdDashto~{f} \rdTo (2,4)<{h} \\
      & & W \SEpbk & \rTo_{c''} & V' \\
      & & \dInto>{h''} & & \dInto>{h'} \\
      & & U & \rTo^{c} & V.
  \end{diagram}
  Considering the lower-left triangle, by
  Lemma~\ref{lemma:decomp:embed}, $f$ is an embedding.  But by
  Lemma~\ref{lemma:one:all}, for any vertex $v_V \in h'(V')$, all of
  $\inv{c} (v_V)$ is in the image of $h$. So, since the pullback is
  isomorphic to $\inv{c} (h' (V'))$, $f$ is surjective on vertices.
  Now, since $c'$ is surjective, each edge $e_V \in h'(V')$ has an
  antecedent $e_{U'}$ in $U'$, hence $f$ is surjective, hence is an
  isomorphism.
\end{proof}

In particular, when $h$ does not cut off any two-ended edge, or
equivalently when $h$ is just a restriction to some of the connected
components of $V$, the process works for any $c$. We call such $h$'s
\emph{cut compatible}.

\paragraph{A sufficient condition for descending plays}
Using Lemmas~\ref{lemma:descent} and~\ref{lemma:descent:embedding}, we
are now able to derive the following sufficient condition by
induction. For $c: U \to V$ any cut only topological play, and $W
\rTo^{r} U$ a play, call \emph{doomed} the edges of $\collapsed{cp}.$
(In the following, we freely write ``doomed in $W$'' when $r$ is clear
from context.)

\begin{definition}
  A play $X \rTo^{p} U$ is \emph{\compatible{$c$}}when for each
  decomposition of $p$ into plays
  \begin{equation}
  \begin{diagram}
    X & \rTo^{q} & W & \rTo^{r} & U,
  \end{diagram}\label{eq:decomp}
\end{equation}
  any edge doomed in $W$, i.e., in $\collapsed{cr}$, has two ends.
\end{definition}

We may characterise \compatible{$c$} plays as follows.
\begin{lemma}\label{lemma:compat}
  A play $p$ is \compatible{$c$} iff for any decomposition 
  \begin{equation}
  \begin{diagram}
    X & \rTo^{q'} & Y & \rInto^{h} & Z & \rTo^{r'} & U
  \end{diagram}\label{eq:decomp:2}
\end{equation}
  of $p$ with $h$ an embedding, $h$ does not cut off doomed edges,
  i.e., if an edge $e \in \collapsed{cr'h}$ is such that $h (e)$ has
  two ends, then $e$ has two ends in $X$.
\end{lemma}

\begin{proof}
  Assume $p$ has a decomposition~(\ref{eq:decomp:2}) as above, but
  with an edge $e \in \collapsed{cr'h}$ lacking at least one end and
  such that $h (e)$ has two ends.  Then, by taking $W = Y$, $q = q'$,
  and $r = r'h$, $e$ contradicts \compatibility{$c$} of $p$.

  Conversely, it is enough to show that for any $X \rTo^{p} U$
  satisfying the condition, any edge in $\collapsed{cp}$ has two ends.
  We proceed by induction on $p$, using
  Lemma~\ref{lemma:descent:embedding} for the induction step (the case
  of \propermoves being easy).
\end{proof}

We have:
\begin{lemma}\label{lemma:proper:descent}
  Any \compatible{$c$} play $p$ descends to some cut free play along
  $c$, in a square
  \begin{equation}
    \begin{diagram}
      X & \rTo^{c'} & Y \\
      \dTo<{p} & & \dTo>{p'} \\
      U & \rTo^{c} & V
    \end{diagram}\label{eq:descend:proper:embeddings}
  \end{equation}
  with $p'$ a cut free play and $c'$ a cut only topological play.
  Again, $\collapsed{cp} = \collapsed{c'}$.
\end{lemma}

  \begin{proof}
    By induction. The induction step uses Lemmas~\ref{lemma:descent},
    \ref{lemma:descent:embedding}, and~\ref{lemma:compat}.
  \end{proof}

  We will now define our $c$-cables using \compatibility{$c$}.

\subsection{Cables}\label{subsec:cables}
Given a cut only topological play $U \rTo^{c} V$ as above, a play $Y
\rTo^{r} U$ is $c$-\emph{atomic} when $Y$ is connected and its edges
are doomed exactly when they have two ends.

\begin{definition}
  A $c$-\emph{cable} is a \compatible{$c$} play $W \rTo^{p} U$ such that
  for any decomposition
  \begin{diagram}
    W & \rTo^{q} & X & \rTo^{m} & Y & \rTo^{r} & U
  \end{diagram}
  of $p$ with $m$ a \propermove, $r$ is $c$-atomic.
\end{definition}

But, we have
\begin{lemma}
  Being atomic is equivalent to being connected and having no
  two-ended edge.
\end{lemma}
\begin{proof}
  Atomic hypersequents satisfy the condition. Conversely, non atomic,
  connected hypersequents all have at least one two-ended edge.
\end{proof}

This yields:
\begin{lemma}\label{lemma:c:atomic}
  A \compatible{$c$} play $Y \rTo^{r} U$ is $c$-atomic iff
  in its descent square
  \begin{diagram}
    Y & \rTo^{c_Y} & Y' \\
    \dTo<{r} & & \dTo>{r'} \\
    U & \rTo^{c} & V,
  \end{diagram}
  $Y'$ is atomic.
\end{lemma}
\begin{proof}
  Let $E$ be the set of two-ended edges in $Y$, not in
  $\collapsed{c_Y}$, i.e., not in $\collapsed{cr} = \collapsed{c_Y}$.
  Let $E'$ be the set of two-ended edges in $Y'$.  Since the edges
  outside $\collapsed{c_Y}$ are left untouched by $c_Y$, $E'$ is
  non-empty iff $E$ is non-empty.

  Moreover, $c_Y$ is a topological play, so $Y$ is connected iff $Y'$
  is connected.

  By the previous Lemma, this gives the expected result.
\end{proof}

Finally, this entails:
\begin{lemma}\label{lemma:cables:descend}
  Any $c$-cable $U' \rTo^{p} U$ descends as a thread.
\end{lemma}
\begin{proof}
  Cables are \compatible{$c$}, so we may consider the descent square
  \begin{diagram}
    U' & \rTo^{c'} & V' \\ 
    \dTo<{p} & & \dTo>{p'} \\
    U &  \rTo^{c} & V.    
  \end{diagram}
  Now, for $p'$ to be a thread, it suffices to consider any of its
  decompositions as
  \begin{diagram}
    V' & \rTo^{q'} & X' & \rTo^{m'} & Y' & \rTo^{r} & V
  \end{diagram}
  and show that $Y'$ is atomic.
  But descent is defined inductively, so such a decomposition yields
  a decomposition 
  \begin{diagram}
    U' & \rTo^{c'} & V' \\ 
    \dTo<{q} & & \dTo>{q'} \\
    X & \rTo^{c_X} & X' \\ 
    \dTo<{m} & & \dTo>{m'} \\
    Y & \rTo^{c_Y} & Y' \\ 
    \dTo<{r} & & \dTo>{r'} \\
    U &  \rTo^{c} & V  
  \end{diagram}
  of the above descent square.  Because $p$ is a $c$-cable, $r$ is
  $c$-atomic, so by the previous lemma, $Y'$ is atomic.
\end{proof}

We now turn to exploiting this to descend strategies.  To do that, we
need to carefully select the $c$-cables complying with a given
strategy. We first define in the next section the threads
$\threadsof{p}$ underlying a given play $p$, and then define the
$c$-cables of a strategy $\proc$ to be those $c$-cables $p$ such that
$\threadsof{p} \subseteq \proc$.

\subsection{The threads of a play}\label{subsec:threads:of:play}

To any play $\play$ on $U$, what its set of threads should be is
intuitively clear, but is a bit tricky to formalise.  What we do is
define a graph $\PlaysHU$ of ``embeddings'' between plays on
$U$. Intuitively, in this graph, an edge $p \to p'$ indicates how at
each stage $p$ sees part of what happens in $p'$. The set
$\threadsof{p}$ of threads of $p$ will then consist of all threads $t$
with an edge $t \to p$.  This extends by union to sets of plays, so,
to any strategy $\proc$ on $U,$ we may associate the set
$\playsof{\proc}$ of plays $\play$ on $U$ whose threads are all in
$\proc$, i.e., such that $\threadsof {\play} \subseteq \proc.$

It remains to define our graph $\PlaysHU$.  First, consider the graph
$\MMA$ with vertices the embeddings $i: U \rInto V$ and whose edges $i
\to j$ have one of the following forms

\begin{mathpar}
  \begin{minipage}[c]{0.4\linewidth}
    \begin{diagram}
      U \SEpbk & \rInto^{i} & V \\
      \dTo<{m'} & & \dTo>{m} \\
      U' & \rInto^{j} & V'
    \end{diagram}
  \end{minipage}
\and
  \begin{minipage}[c]{0.4\linewidth}
    \begin{diagram}
      U  & \rInto^{i} & V \\
      & \rdInto<{j} & \dTo>{m} \\
      & & V'
    \end{diagram}
  \end{minipage}
\and \\
  \begin{minipage}[c]{0.4\linewidth}
    \begin{diagram}
      U & \rInto^{i} & V \\
      & \rdInto<{j} & \dInto>{k} \\
      & & V'
    \end{diagram}
  \end{minipage}
\and
  \begin{minipage}[c]{0.4\linewidth}
    \begin{diagram}
      U & \rInto^{i} & V \\
      \dInto<{h} & \ruInto<{j} & \\
      U', & & 
    \end{diagram}
  \end{minipage}
\end{mathpar}
where the first square is a pullback and the second diagram is such
that the induced square
\begin{diagram}
  U \SEpbk & \rInto^{i} & V \\
  \dEqto & & \dTo>{m} \\
  U & \rInto^{j} & V'
\end{diagram}
is a pullback, and where $m$ and $m'$ denote \propermoves and $h$ and
$k$ denote embeddings, all seen as \generalizedmoves.  This graph
freely generates a category $\PPA$, whose morphisms are piles of such
diagrams.  Furthermore, there are morphisms of graphs $s, t: \MMA \to
\PA$ sending the squares and triangles to their vertical borders. By
adjunction, they induce functors $s, t: \PPA \to \PA$. This structure
now induces a ``horizontal'' graph $\PlaysH$ whose vertices are plays,
and whose edges $p \to q$ are morphisms $i \to j$ in $\PPA$ with
left-hand border $p$ and right-hand border $q$.

Finally, the graph $\PlaysHU$ evoked above has vertices the plays on
$U$, and edges the edges in $\PlaysH$ with lower border the identity.
Thus, $\threadsof{\play}$ is the set of threads $t$ on $U$ such that
there exists an edge $t \to p$ in $\PlaysH$ with lower border the
identity.

\subsection{Cut elimination}\label{subsec:cutelim}
For any strategy $\proc$ on $U$, we at last define the set
$\pipesofc{\proc}$ of $c$-\emph{cables} of $\proc$ to be the set of
$c$-cables $p$ with $\threadsof{p} \subseteq \proc$. 

Recall that $\descend_c$ sends sets of plays $S$ to the set of plays
descending from plays in $S$ along $c$. We set:
\begin{definition}
  For any strategy $\proc$, let $$\celimc{\proc} =
  \descend_c(\pipesofc{\proc}).$$
\end{definition}
We then obtain:
\begin{lemma}\label{lemma:quasi:strats}
  The set of threads $\celimc{\proc}$ satisfies axioms \textbf{S1} to
  \textbf{S3} for strategies, plus one direction of axiom \textbf{S4},
  namely that (in the same setting) if $t \rond \geom(hh') \rond t'
  \in \proc$, then $t \rond h \rond h' \rond t' \in \proc$.
\end{lemma}
We first prove that the corresponding direction of \textbf{S4} holds
for cables:
\begin{lemma}
  For any strategy $\proc$, if $t \rond \geom(hh') \rond t' \in
  \pipesof{\proc}$, then $t \rond h \rond h' \rond t' \in
  \pipesof{\proc}$.
\end{lemma}
\begin{proof}
  Being a thread of a cable is insensitive to composition or 
  decomposition of embeddings. Indeed, any edge in the graph $\PlaysHU$
  between sequences of embeddings may be obtained by piling up
  triangles as in
  
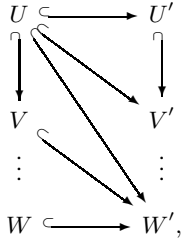
\begin{diagram}
    U & \rInto & U' \\
    \dInto & \rdInto \rdInto (2,4) & \dInto \\
     V  &&   V' \\
     \vdots  & \rdInto &   \vdots \\
     W & \rInto &   W',
  \end{diagram}
  with no constraint on the numbers of embeddings on each side; only
  the commutativity of the outer diagram matters in the end.
\end{proof}

\begin{proof}[Proof of Lemma~\ref{lemma:quasi:strats}]
  By Lemma~\ref{lemma:cables:descend}, $\celimc{\proc}$ is a
  (non-empty) set of threads. Also, since $\proc$ is a strategy,
  $\celimc{\proc}$ is prefix-closed.  Furthermore, for any mandatory
  \generalizedmove $f$ extending $p' \in \celimc{\proc}$, $f$ easily
  lifts to a mandatory move in the corresponding cable, which descends
  as $f$, hence $\celimc{\proc}$ is stable under extension by
  mandatory moves.  Furthermore, $\celimc{\proc}$ is stable under
  isomorphism, by construction of $\descend$. Finally, if $t \rond
  \geom(hh') \rond t'$ is in $\celimc{\proc}$, then $\geom (hh')$
  comes from an edge in $\descend$, i.e., a square
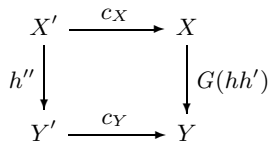
  \begin{diagram}
    X' & \rTo^{c_X} & X \\
    \dTo<{h''} & & \dTo>{\geom (hh')} \\
    Y' & \rTo^{c_Y} & Y
  \end{diagram}
  with $h''$ an embedding. But by the pullback Lemma and
  Lemma~\ref{lemma:descent:embedding}, we may choose pullbacks as in
  
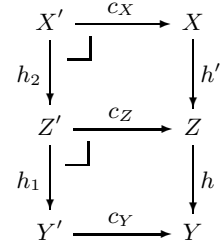
\begin{diagram}
    X' \SEpbk & \rTo^{c_X} & X \\
    \dTo<{h_2} & & \dTo>{h'} \\
    Z' \SEpbk & \rTo^{c_Z} & Z \\
    \dTo<{h_1} & & \dTo>{h} \\
    Y' & \rTo^{c_Y} & Y
  \end{diagram}
  such that $h_1 h_2 = h''$. But then by \textbf{S4}
  for $\proc$, we could replace $h''$ by $h_1 h_2$ in the cable
  descending to $t \rond \geom(hh') \rond t'$, and obtain a cable
  descending to $t h h' t'$.

\end{proof}

However, as we have seen in Section~\ref{subsec:celim:overview},
the set of threads $\celimc{\proc}$ need not be closed under
the other direction of \textbf{S4}.
But we may perfectly close a set of threads under composition of
embeddings: consider the rewriting relation on plays defined by
$$t h h' t' \rightarrow t \rond \geom (h h') \rond t,$$
and given a set of threads $\proc$, let $\closure{\proc}$
be the set of plays reachable from $\proc$ by this relation.
We have
\begin{lemma}
  The set of threads $\closure{\celimc{\proc}}$ is a strategy.
\end{lemma}
\begin{proof}
  Axioms \textbf{S1}-\textbf{S3} are preserved by $\closurefun$, as
  well as the first direction of Axiom \textbf{S4}. The second
  condition is now satisfied, hence $\closure{\celimc{\proc}}$ is a
  strategy.
\end{proof}

We may then define our family of functions: for any cut only
topological play $U \rTo^{c} V$ and set of threads $\proc$ on $U$, let
$\cutelimc{\proc} = \closure{\celimc{\proc}}$, and
$\cutelimalong{U}{\proc} = \closure{\celimalong{\id_U}{\proc}}$.  We
have seen that if $\proc$ is a strategy, then so is
$\cutelimalong{U}{\proc}$. We further have
\begin{lemma}
  The functions $\cutelimfun: \stratsH (U) \to \cfH (U)$
  define a morphism of sheaves.
\end{lemma}
\begin{proof}
Restriction commutes with cut elimination.  
\end{proof}

\section{Logic}\label{sec:logic}
We at last start using our game as a model of MALL.  We first define
winning strategies, and we relate them to more standard notions, and
discuss categories of games and strategies.  We then show that winning
strategies are stable under cut elimination, and obtain coherence as a
corollary. We then show that every MALL proof generates a winning
strategy and \emph{vice versa}, hence our model is correct and
complete. (This would have entailed coherence in a less direct way.)

\subsection{Winning strategies}\label{subsec:winning}
When should a strategy be winning? Since at any stage and on any
sequent it has to accept all negative moves, it reaches sequents with
only $\TT$'s and positive edges. In such a sequent, if there actually
are some $\TT$ edges, then the play should be considered won, thanks
to the $\TT$ axiom of MALL. Otherwise, there are only positive edges,
and the strategy should propose a positive \propermove.  In other
words, when a winning strategy is stuck, it has to be on a position
with a $\TT$ edge.

More formally:
\begin{definition}
  A sequent is \emph{positive} when it has no input edge.  A set of
  threads $\proc$ is \emph{winning} when it is non empty, stable under
  extension by mandatory \generalizedmoves, and when every thread in
  $\proc$ ending on a positive sequent has an extension by a
  \propermove in $\proc$.
\end{definition}

Equivalently, we may call a play $V \rTo^{p} U$ \emph{maximal} in some
set of threads $\proc$ if it is atomic and has no extension by a
\propermove in $\proc$.  If $\proc$ is a strategy, then for such a
maximal $p$, any negative edge of $V$ is labeled $\TT$, otherwise
there is an extension by a passive \propermove.  Let now such a
maximal play in a set of threads $\proc$ be \emph{won} when $V$ is
either empty, or a ($\TT$) edge, or a sequent with a negative edge.
Otherwise it is \emph{lost}. A position is maximal if it is the domain
of a maximal thread.

\begin{lemma}
  A strategy is \emph{winning} iff it is non-empty and all its
  maximal positions are won.
\end{lemma}

In game theory, and in particular in game semantics, strategies are
usually defined as sets of plays (without embeddings).  This raises
the question: in which sense is the notion of a winning strategy
$\proc$ related to its set of plays $\playsof{\proc}$?

First, observe that the set of plays $\playsof{\proc}$ is
prefix-closed, and \emph{welcoming}, in the sense that it is stable
under extension by a passive \propermove or an embedding. Indeed, the
threads of any such extension of a play $p \in \playsof{\proc}$ are
either already threads of $p$, or extensions of one of them by a
passive \propermove or an embedding, hence again in $\playsof{\proc}$.

Let a play $p$ in a set of plays $P$ be \emph{maximal} when $p$ has no
extension by a \propermove in $P$. Call a maximal $p$ \emph{won} when
all its sequents have at least one negative edge. Observe that it is
different to be maximal as a thread or as a play: a thread is maximal
as a thread only if its domain $V$ is atomic.

\begin{lemma}
  If $\proc$ is a winning strategy, then any maximal play in
  $\playsof{\proc}$ is won.
\end{lemma}
\begin{proof}
  Assume given a play $p: V \to U$ in $\playsof{\proc}$ with a sequent
  $s$ without any negative edge, and consider a thread $t$ leading to
  it in $\proc$. If $s$ has no edge, then $t$ is maximal and lost,
  contradicting the winning character of $\proc$. Thus, $s$ has some
  positive edges.  But again, since $\proc$ is winning, $t$ cannot be
  maximal, so it has an extension by a \propermove. Other sequents have to
  accept this move because $\proc$ is winning, so $V$ could not be
  maximal.
\end{proof}

All in all, we have
\begin{theorem}\label{theorem:plays}
  The set of plays $\playsof{\proc}$ of a winning strategy is
  non-empty, prefix-closed, and welcoming, and its maximal plays are
  all won.
\end{theorem}

However, a set of plays may satisfy the conditions of
Theorem~\ref{theorem:plays} without being generated by a strategy. The
main reason is because these conditions miss stability under
restriction and amalgamation. For instance, on the hypersequent
\grafflecenter{cex} consider the winning set of plays $P$ with
\begin{itemize}\item 
  proper plays choosing one repartition of the left-hand $\un$ edges,
\item plays after restriction to the left-hand sequent choosing the
  other,
\end{itemize}
which satisfies the conditions, but is not generated by a strategy.
Indeed, any strategy having at least the threads in $\threadsof{P}$
would allow both repartitions of the left-hand $\un$ edges in its
global plays.

\subsection{Coherence, correctness, and completeness}
We now turn to proving the announced logical results: coherence,
correctness, and completeness. We start by proving that winning
strategies are stable under cut elimination.
\begin{theorem}
  If $\proc$ is winning, then so is $\cutelimc{\proc}$.
\end{theorem}
\begin{proof}
  First, if $\celimc{\proc}$ is winning, then so is
  $\closure{\celimc{\proc}}$, since no maximal positions are
  added. Let us thus show that $\celimc{\proc}$ is winning.

  Let $p': V'' \to V$ lead to a maximal position $V''$ in
  $\celimc{\proc}$, and choose $p: U'' \to U$ as in
  \begin{diagram}
    U'' & \rTo^{c''} & V'' \\
    \dTo<{p} && \dTo>{p'} \\
    U & \rTo^{c} & V
  \end{diagram}
  maximal in $\pipesofc{\proc}$ descending to $p'$, i.e., $p$ has no
  extension in $\pipesofc{\proc}$ also descending to $p'$. By
  maximality, $V''$ is atomic, so by Lemma~\ref{lemma:c:atomic}, $p$
  is $c$-atomic.  So, if $V''$ is either an edge or empty, then so is
  $U''$.  Otherwise, $V''$ is a sequent, so $U''$ is a connected
  hypersequent with the same one-ended edges. Now, we claim that every
  sequent in $U''$ has a negative edge. Indeed, if any sequent there
  had
  \begin{itemize}
  \item no edge at all, then a thread in $\proc$ would lead to it,
    contradicting the winning character of $\proc$,
  \item only positive edges, then because $\proc$ is a winning strategy
    $U''$ would admit an extension by a \propermove in $\pipesofc{\proc}$,
    contradicting its maximality.
  \end{itemize}
  So, each sequent in $U''$ has at least one negative edge.  But this
  easily implies that $U''$ has at least one input edge, which then
  has to be a $\TT$. Therefore, $V''$ has an input $\TT$ edge and is
  thus won.
\end{proof}

This directly entails coherence:
\begin{corollary}
  There is no winning strategy on the empty sequent.
\end{corollary}
\begin{proof}
  Any winning strategy $\proc$ would yield a cut free one
  $\cutelim{\proc}$. But the latter cannot be winning, as it has no
  \propermove, so the empty sequent is maximal, but lost.
\end{proof}

\subsection{Correctness and completeness}
We now investigate the correspondence with provability in MALL. We
defer a proof theoretical investigation to further work.

\begin{lemma}
  If a sequent $\Gam$ is provable in MALL, then there is a winning
  strategy on it.
\end{lemma}

\begin{proof}
  By standard proof technology, $\Gam$ admits a cut free proof $\pf$
  which at any stage starts by completely breaking negative
  connectives, and with no axiom links (i.e., conclude by $\un$ and
  $\TT$ rules). 

  We proceed by induction on this $\pf$. Observe that at any stage, we
  must accept all embeddings. We do so implicitly, and need only
  specify a strategy when such embeddings lead to a sequent (on edges,
  a strategy has to accept all \generalizedmoves).
  
  Now, let us review the base cases. If $\pf$ is a $\un$ rule, then
  apply the $\un$ move to reach an empty position, which is hence
  won. If $\pf$ is a $\TT$ rule, then by hypothesis the only negative
  formulae of $\Gam$ are $\TT$'s, so there are no possible passive
  \propermoves, and $\Gam$ is maximal, hence won.
  
  For the induction step, first, accept all passive \propermoves, which in
  various paths lead to a sequent $\Gam'$ with only $\TT$'s and
  positive formulae.  The proof $\pf$ chooses one such path to
  $\Gam'$. Now, if this path is non empty, then the size has
  decreased, so we may apply the induction hypothesis.  Otherwise,
  $\Gam = \Gam'$, and $\pf$ starts with an active \propermove $m$, reaching
  premisses $\pf_1, \ldots, \pf_n$, with $n \in \ens{1, 2}$, which in
  turn have to perform negative rules to reach premisses $\pf'_1,
  \ldots, \pf'_n$.  We choose $m$ as the next move of our
  strategy. Then, accept all passive \propermoves and embeddings, which
  (among others) lead in various paths to the conclusions of $\pf'_1,
  \ldots, \pf'_n$.  Finally, conclude by induction hypothesis.
\end{proof}

\begin{lemma}
  If a sequent $\Gam$ admits a winning strategy, then it is provable in
  MALL.
\end{lemma}
\begin{proof}
  Assume given such a winning strategy $\proc$, which we may suppose
  cut free w.l.o.g.  Since in the game, a play has a finite number of
  \propermoves, we may take this as its size.  The size of a strategy is then
  the maximum size of its plays. 

  Proceeding by induction on the size of $\proc$, if $\proc$ has size
  $0$, then because it is winning on $\Gam$ and $\Gam$ is atomic,
  $\Gam$ is maximal, so it has a $\TT$ edge, hence is an axiom of
  MALL. 

  Otherwise, choose a thread in $\proc$ performing all possible
  passive \propermoves, and reach a sequent $\Gam'$ with only $\TT$'s and
  positive edges.  If the followed path has at least one \propermove, then
  the size has decreased so by induction hypothesis we get a proof of
  $\Gam'$, to which we apply all the corresponding negative rules to
  get a proof of $\Gam$.  Otherwise the followed path is empty, and
  $\Gam = \Gam'$ has only $\TT$'s and positive edges. If it has a
  $\TT$, then $\Gam$ is an axiom of MALL.  

  Otherwise, since $\proc$ is winning, there is a (active) \propermove from
  $\Gam$.
  \begin{itemize}
  \item If it is a $\un$ move, then $\Gam$ is the sequent with 
    exactly one $\un$ formula, which is an axiom of MALL.
  \item If it is a $\plus$ move, then it leads to some sequent $\Gam''$.
    By induction hypothesis, the sequel of $\proc$ being winning, we get
    a proof of $\Gam''$, to which we apply the corresponding rule to get
    a proof of $\Gam$.
  \item If it is a $\tens$ move, then it leads to some hypersequent of
    the shape \grafflecenter{tens6} that is, a disjoint union of two
    sequents.  Since $\proc$ is a strategy, we may follow the restrictions
    to each sequent $\Gam_1$ and $\Gam_2$, apply the induction
    hypothesis there to get proofs $\pf_1$ and $\pf_2$, to which we
    apply the tensor rule to get a proof of $\Gam$. 
  \end{itemize}
\end{proof}

We have proved:
\begin{theorem}
  The topological game for MALL is (logically) sound and complete.
\end{theorem}

\subsection{Towards categories of strategies}\label{subsec:cat}
Without cut elimination, we may construct a category $\prestratsHcat$
of strategies for our topological game for MALL, which we define to be
the strictification of the bicategory of cospans, with
\begin{itemize}
\item objects the formulae, and
\item morphisms $A \to B$ consisting of a cospan
  \begin{diagram}
    A & \rInto & U & \lInto & B
  \end{diagram}
  in $\HH$, equipped with a strategy on $U$, with $A$ and $B$ dangling
  edges labeled with formulae $A$ and $B$, and $U$ a connected
  hypersequent with exactly one input edge -- the image of $A$, and
  one output edge, the image of $B$.
\end{itemize}
Gluing two hypersequents along a edge which is input on one side and
output on the other clearly preserves acyclicity, hence we may hope to
define composition as a strategy on the (chosen) pushout. Now, observe
that the unique strategy on an edge (alone) is the total one, i.e.,
the set of all plays.  Indeed, all \propermoves are passive and atomic. Thus,
any two strategies $p: A \to B$ and $q: B \to C$, have the same
restriction to $B$, hence have a unique amalgamation, which we elect
to be their composition $q \rond p: A \to C$. The identity on $A$ is
given by the unique strategy on the edge labeled $A$. Since winning
strategies are stable under amalgamation, we may form the subcategory
$\prewstratsHcat$ of winning strategies. We could also do the same
with cut free strategies.

However, these categories $\prestratsHcat$ and $\prewstratsHcat$ are
not quite what game semanticists are used to.  Indeed, given two
strategies $\strat: A \to B$ and $\strat': B \to C$, i.e., on objects
like
\begin{center}
  \grafflemininline{.4\linewidth}{typestrat1} \hfil and \hfil
  \grafflemininline{.4\linewidth}{typestrat2} 
\end{center}
respectively, a game semanticist expects their composition
\begin{equation}
\begin{diagram}
  A & \rTo^{\strat} & B & \rTo^{\strat'} & C
\end{diagram}\label{eq:compo:strats}
\end{equation}
to be a strategy on the hypersequent $U$: 
\begin{equation}
\grafflemininline{.5\linewidth}{typestrat4}\label{eq:U}
\end{equation}
not on $V$: 
\begin{equation}
\grafflemininline{.8\linewidth}{typestrat3}\label{eq:V}
\end{equation}
as in $\prestratsHcat$.  

Now, let $\Gam = (A_1, \ldots, A_n)$ and $\Del = (B_1, \ldots, B_m)$
be lists of formulae. We write $U: (\Gam \hyperdash \Del)$ when the
connected hypersequent $U$ has exactly $n$ input edges labeled with
the $A_i$'s and $m$ output edges labeled with the $B_j$'s. For any
such hypersequent $U$, there is a cut only topological play
\begin{diagram}[width=1.5cm]
  (U: \Gam \hyperdash \Del) & \rTo^{c_U} & (\Gam \vdash \Del),
\end{diagram}
Thus, for any (winning) strategy $\proc$ on $U$, there is a (winning)
strategy $\cutelimalong{c_U}{\proc}$ on $\Gam \vdash \Del$.  In order
to obtain categories of strategies closer to usual game semantics, we
might want to quotient our categories $\prestratsHcat$ and
$\prewstratsHcat$ by decreeing that two strategies $(U, \proc)$ and
$(V, \procc)$ from $A$ to $B$ are equivalent when
$\cutelimalong{c_U}{\proc} = \cutelimalong{c_V} {\procc}$.
Alternatively, we could take morphisms $A \to B$ to be (winning)
strategies on the sequent $A \vdash B$, and composition to be defined
by amalgamation followed by descent along the cut play, say,
from~(\ref{eq:V}) to~(\ref{eq:U}) above.  However, this appears
trickier than expected, specifically w.r.t.\ associativity of
composition in the obtained candidate categories, and we leave it for
further work.

\section{Related and further work}
The game in this paper is almost the same as in an earlier talk
\citep{H3:huet}, with a few evolutions. The development is very
different: in \citet{H3:huet}, we were concerned with making plays a
stack, which here is avoided by passing directly to strategies.  The
notion of strategy we adopt here is radically new -- \citet{H3:huet}
used sets of proper plays. Finally, we provide correctness and
completeness results which were not in \citet{H3:huet}.

\citet{Miller08} investigate a closely related game, with analogous
correctness and completeness results.  Their approach first
technically differs in the way the game ends, and in the definition of
won positions.  More importantly, their game does not feature any cut
move, so they do not deal with cut elimination in any sense.  Finally,
they do not use topological methods at all.

\citet{Mellies04} and subsequent papers propose notions of
games where plays should be considered up to permutation of certain
moves. Our game certainly has an asynchronous flavor in this sense,
where permutations arise directly from the topology. However, a formal
connection remains to be established.

Most striking are probably the similarities with \citegenitif{locus}
ludics, from which we gratefully acknowledge inspiration. A first
difference is technical: Girard does not use topological methods at
all, maybe because ludics are restricted to a very particular form of
graphs. Also, our game is closer to MALL sequent calculus than ludics,
e.g., it does not feature the \emph{daimon} move of ludics.
Furthermore, our edges are labeled with formulae, which fixes their
behavior -- there is exactly one strategy per edge. A key ingredient
to Girard's approach is to avoid labels, and instead say that a
strategy follows a typing (i.e., a labeling of edges with formulae)
when its restriction to each edge behaves accordingly. Adapting our
game to this approach is left for further work. Finally, ludics'
strategies are still defined as sets of plays, i.e., non locally.

More intrinsically to our game, there are a number of possible
directions for improvement. First, as evoked in
Section~\ref{subsec:cat}, our game has to be adapted to fit into a
category of strategies. Furthermore, one might want to tighten the
connection between strategies and proofs, e.g., towards a full
completeness result.  Finally, we will try to extend our game to
exponentials, and (at least first-order) quantification. This promises
to be more difficult, particularly w.r.t.\ noetherianness.

\bibliographystyle{plainnat} \bibliography{b}

\newcommand{\online}[1]{Available at \texttt{#1}\ensuremath{\;}}
\begin{thebibliography}{7}
\expandafter\ifx\csname natexlab\endcsname\relax\def\natexlab#1{#1}\fi
\expandafter\ifx\csname url\endcsname\relax
  \def\url#1{{\tt #1}}\fi

\bibitem[Abramsky(1997)]{abramsemint}
Samson Abramsky.
\newblock {\em Semantics and Logics of Computation}, chapter Semantics of
  interaction: an introduction to game semantics.
\newblock Cambridge University Press, 1997.

\bibitem[Delande and Miller(2008)]{Miller08}
Olivier Delande and Dale Miller.
\newblock A neutral approach to proof and refutation in {MALL}.
\newblock In {\em LICS '08}, 2008.
\newblock To appear.

\bibitem[Girard(1987)]{ll}
Jean-Yves Girard.
\newblock Linear logic.
\newblock {\em Theoretical Computer Science}, 50:\penalty0 1--102, 1987.

\bibitem[Girard(2001)]{locus}
Jean-Yves Girard.
\newblock Locus solum: From the rules of logic to the logic of rules.
\newblock {\em Mathematical Structures in Computer Science}, 11\penalty0
  (3):\penalty0 301--506, 2001.

\bibitem[Hirschowitz et~al.(2007)Hirschowitz, Hirschowitz, and
  Hirschowitz]{H3:huet}
Andr\'e Hirschowitz, Michel Hirschowitz, and Tom Hirschowitz.
\newblock Playing for truth on mutable graphs.
\newblock Talk at the Colloquium in honor of G\'erard Huet's 60th birthday,
  ENS, Paris, 2007.

\bibitem[Hyland(1997)]{hyland97}
Martin Hyland.
\newblock {\em Semantics and Logics of Computation}, chapter Game Semantics.
\newblock Cambridge University Press, 1997.

\bibitem[Melli{\`e}s(2004)]{Mellies04}
Paul-Andr{\'e} Melli{\`e}s.
\newblock Asynchronous games 2: the true concurrency of innocence.
\newblock In {\em Concurrency Theory '04}, volume 3170 of {\em Lecture Notes in
  Computer Science}, pages 448--465. Springer, 2004.

\end{thebibliography}

\acks{The third author acknowledges support from ANR projects
  MoDyFiable and Choco. We also acknowledge useful feedback from
  Florian Hatat, who spotted quite a few errors, and Fran\c{c}ois
  M\'etayer.}
\end{document}